\documentstyle[12pt,epsf]{article} 
\def\gsim{\mathrel{\rlap {\raise.5ex\hbox{$ > $}}
{\lower.5ex\hbox{$\sim$}}}}
\def\lsim{\mathrel{\rlap {\raise.5ex\hbox{$ < $}}
{\lower.5ex\hbox{$\sim$}}}}

\newcommand{\be}{\begin{equation}}
\newcommand{\ee}{\end{equation}}
\newcommand{\bea}{\begin{eqnarray}}
\newcommand{\nn}{\nonumber}
\newcommand{\eea}{\end{eqnarray}}

\def\gappeq{\mathrel{\rlap {\raise.5ex\hbox{$>$}}
{\lower.5ex\hbox{$\sim$}}}}
 
\def\lappeq{\mathrel{\rlap{\raise.5ex\hbox{$<$}}
{\lower.5ex\hbox{$\sim$}}}}

\begin{document} 
\begin{titlepage} 

\begin{flushright} 
OUTP-99-32P \\
cond-mat/9909310
\end{flushright} 

\vspace{0.1in} 
\begin{centering} 

{\Large {\bf  Comments on Extended t-J Models, Nodal Liquids and 
Supersymmetry$^{\diamond}$}} \\
\vspace{0.2in} 
{\bf Nick E. Mavromatos }$^{*}$ \\
\vspace{0.1in} 
Department of Physics (Theoretical Physics), University of Oxford \\
1 Keble Road, Oxford OX1 3NP, U.K. and \\
CERN, Theory Division, Geneva 23 CH-1211, Switzerland \\

\vspace{0.2in} 
{\bf Sarben Sarkar} \\
\vspace{0.1in} 
Wheatstone Physics Laboratory, King's College London, \\
Strand, London WC2R 2LS, U.K.  \\

\vspace{0.4in} 
{\bf Abstract}

\end{centering} 

{\small In the context of extended $t-J$ models, with intersite Coulomb interactions of the form
$-V\sum\limits_{\left\langle {i,j} \right\rangle } {n_in_j}$, with $n_i$ denoting the  electron number
 operator at site $i$, nodal liquids are discussed. 
We use the spin-charge separation ansatz as applied
to the nodes of a d-wave superconducting gap.
Such a situation may be of relevance to 
the physics of high-temperature 
superconductivity. 
We point out the 
possibility that at certain points of the parameter space 
supersymmetric
points may occur, characterized by dynamical supersymmetries 
between the spinon and holon degrees of freedom, 
which 
are quite different from the symmetries in conventional 
supersymmetric $t-J$ models. 
Such symmetries pertain to the continuum 
effective field theory of the nodal liquid, and 
one's hope is that the ancestor lattice model 
may differ from the continuum theory only by 
renormalization-group irrelevant operators in the infrared. 
We give plausible arguments that nodal liquids at 
such supersymmetric points are 
characterized 
by superconductivity of Kosterlitz-Thouless
type. 
The fact that 
quantum fluctuations around such points 
can be studied in a 
controlled way, probably makes such systems 
of special importance 
for an eventual non-perturbative understanding
of the complex phase diagram of the associated 
high-temperature superconducting 
materials.}

\vspace{0.90in}
\begin{flushleft}  
$^{\diamond}$ Presented by N.E.M. at the Workshop {\it Common Trends 
in Particle and Condensed Matter Physics}, September 24-28 1999, Corfu 
(Greece), adjacent to the TMR meeting {\it Topology and Phase Transitions
in Hot Matter}, September 19-26 1999. \\
$^{*}$ {\it Address after Oct. 1 1999}~: 
Wheatstone Physics Laboratory, King's College London, 
Strand, London WC2R 2LS, U.K. \\
\end{flushleft} 

\end{titlepage}

\section{Introduction}

The study of strongly correlated electron systems (SCES) is a major 
enterprise in modern condensed matter physics primarily due to high temperature
(planar) superconductors, fractional Hall conductors
and more recently in semiconductor quantum dots.
Owing to various non-Fermi liquid features of SCES many believe that the
 low-energy excitations of 
these systems are influenced by the proximity of a critical Hamiltonian
in a generalized coupling-constant space. In this scenario,
known as spin-charge separation~\cite{anderson},
these excitations are spinons, holons and gauge fields.

Important paradigm for SCES are the conventional Hubbard model, 
or its $t-j$ extension, both of which have been conjectured
to describe the physics of high-temperature 
superconducting doped antiferromagnets. 
Numerical simulations of such models~\cite{dagotto}, 
in the presence of
very-low doping, have provided evidence for electron substructure (spin-charge
separation) in such systems.

In ref. \cite{[5]},  
an extension of the spin-charge separation
ansatz, allowing
for a particle-hole symmetric formulation away from half-filling,
was introduced by writing:
\be
\chi _{\alpha \beta }\equiv \left( {\matrix{{\psi _1}&{\psi _2}\cr
{-\psi _2^\dagger}&{\psi _1^\dagger}\cr
}} \right)_i\left( {\matrix{{z_1}&{-\bar z_2}\cr
{z_2}&{\bar z_1}\cr
}} \right)_i\,,\label{1.5}
\ee
where
the fields $z_{\alpha,i}$ obey canonical {\it bosonic}
commutation relations, and are associated with the
{\it spin} degrees of freedom (`spinons'), whilst the fields $\psi$
are Grassmann variables,obeying Fermi statistics, and are associated with the electric charge degrees of freedom
(`holons').
There is a hidden non-abelian gauge symmetry~
$SU(2) \otimes U_S(1)$ in the ansatz,
which becomes a dynamical symmetry
of the pertinent planar Hubbard model, studied in ref. \cite{[5]}. 

The ansatz (\ref{1.5}) is different from that of refs.~\cite{[17]},
where the holons are represented as charged bosons, and the spinons as
fermions. That framework, unlike ours, is not a convenient starting point
for making predictions such as the behaviour of the
system under the influence of strong external fields.
As argued in \cite{fmijmpb}, 
a strong magnetic field induces the opening of a second
 superconducting gap at the nodes of the $d$-wave gap, in agreement
with recent experimental findings on the behaviour of the thermal conductivity
of high-temperature cuprates 
under the influence of strong external magnetic fields~\cite{ong}.

In \cite{[5]} a single-band 
Hubbard model was used. Such a model 
should not be regarded as merely phenomenological
for cuprate superconductors in the sense that it can be rigorously derived
from chemically realistic multiband models with extra
nearest-neighbour interactions of the form~\cite{[1]}:
\be
H_{int} = -V \sum_{<ij>}
\eta_i \eta_j \qquad \eta_i \equiv \sum_{\alpha=1}^{2}
c_{\alpha,i}^\dagger c_{\alpha,i}\,,\label{1.7}
\label{extraint}
\ee

What we shall argue below is that the presence of 
interactions of the form (\ref{extraint}) is crucial 
for the appearence of supersymmetric points
in the parameter space of the spin-charge 
separated model. Such points 
occur for particular doping concentrations.
In this talk we shall only sketch the basic ideas.
A more detailed account of the work will be given  
in a future publication~\cite{future}.
As we shall discuss, 
this supersymmetry 
is a {\it dynamical symmetry}
of the spin-charge separation, 
and occurs between 
the spinon and holon degrees of freedom 
of the ansatz (\ref{1.5}). 
Its appearance may indicate the onset of 
unconventional 
superconductivity of Kosterlitz-Thouless (KT) type~\cite{[2],[3]} 
in the liquid of excitations about the nodes of the 
d-wave superconducting gap
(``nodal liquid''), to which we restrict our attention
for the purposes of this work. 

It should be stressed that 
the supersymmetry characterizes the 
continuum {\it relativistic} effective (gauge) 
field theory of the nodal 
liquid. The ancestor lattice model is of course
{\it not supersymmetric} in general. What, however, one hopes
is that at such supersymmetric points the 
universality 
class of the continuum low-energy 
theory is the {\it same} 
as that of the lattice model, in the sense that the latter
differs from the continuum effective theory only by 
renormalization-group {\it irrelevant} operators (in the infrared). 
This remains to be checked by detailed
studies, which do not constitute the topic of this talk. 

Supersymmetry 
provides, in general, a much more 
controlled way for 
dealing with quantum fluctuations about
the ground state of a field-theoretic system
than a non-supersymmetric theory~\cite{[14]}. 
In this sense, one hopes that by working in such 
supersymmetric points in the parameter space of the 
nodal liquid she/he might obtain some exact results 
about the phase structure, which might be useful 
for a non-perturbative understanding of the 
complex phase diagrams that characterize the 
physics of the (superconducitng) doped antiferromagnets. 

Significant progress towards a non-perturbative 
understanding of Non-Abelian gauge field theories 
based on supersymmetry have been made 
by Seiberg and Witten~\cite{[15]}.
The fact that the spin-charge separation 
ansatz (\ref{1.5}) of the doped antiferromagnet is known to be characterized
by such non-Abelian gauge  structure is an encouraging sign. 
However, it should be noted that in the case of ref. \cite{[15]}
extended supersymmetries were necessary for yielding exact results.
As we shall discuss below, in the case of doped 
antiferromagnets,
and under special conditions,
the supersymmetric points are characterized by $N=1$ 
three-dimensional supersymmetries, although 
under certain circumstances the supersymmetry may be elevated
to $N=2$~~\cite{[8]}, for which 
some exact results concerning the phase structure
can be obtained~\cite{[16]}. 
However, in the realistic circumstances
of a condensed-matter system such as a high-temperature
superconductor, 
even the $N=1$ supersymmetry
of the supersymmetric points is expected to be broken 
at finite temperatures or under the influence of external 
elctromagnetic fields.  
Nevertheless, one may hope that by viewing 
the case of broken supersymmetry as the result 
of some perturbation that takes one 
away from the supersymmetric point,
valuable non perturbative information may still be obtained.
As we shall see, a possible example of this cocnerns 
the above-mentioned KT superconducting properties~\cite{[2]} 
that characterize such points. 

The structure of the talk is as follows: 
In section 2 we describe briefly 
the statistical model
which gives rise to the continuum relativistic 
effective (2+1)-dimensional field theory of the nodal liquid.
In section 3 we discuss the properties and
(non-abelian gauge) symmetries  of the spin-charge separation
ansatz that characterizes the model. 
In the next section we discuss the intersite Coulomb 
interactions, which are of crucial importance for the 
existence of supersymmetric points. In section 5 we 
state the conditions for N=1 supersymmetry at such points,
and describe briefly their importance for 
yielding superconductivity of Kosterlitz-Thouless type. 
We conclude in section 6 with some prospects for future 
work.

\section{The Model and its Parameters}

In reference \cite{[1]} 
it was argued that BCS-like scenarios for high $T_c$ superconductivity based on
 extended $t-J$ models yield reasonable predictions for the critical temperature $T_c^{\max }$ at optimum doping . 
There it was argued that a pivotal role was played by next-to-nearest neighbour and third neighbour hoppings, ${t'}$ 
and ${t'''}$ respectively. In particular the combination $t_-\equiv t'-2t'''$ determines the shape of the Fermi surface 
and the nature of the saddle points and the associated $T_c^{\max }$.

Our aim is to use the extended $t-J$ model studied  
in \cite{[1]} in order to discuss the appearance of relativistic 
charge liquids at the nodes of the associated d-wave superconducting gap. We will argue that the nodes characterize
the model in a certain range of parameters. 
We will demonstrate that at a certain 
regime of the parameters and doping 
concentration the nodal liquid effective 
field theory of spin-charge separation 
exhibits supersymmetry. This supersymmetry is dynamical and should not be confused with the non-dynamical symmetry
under a graded supersymmetry algebra that characterises the spectrum of doped antiferromagnets at two special points of
the parameter space \cite{[4]}. We shall also 
discuss unconventional mechanisms for superconductivity
in the nodal liquid similar to 
the ones proposed in \cite{[2],[3]}.

To start with let us describe briefly 
the extended $t-J$ model used in Ref. \cite{[1]}. 
The Hamiltonian is given by:
\be
H=P\left( {H_{hop}+H_J+H_V} \right)P + PH_\mu P\,,\label{1.1}
\ee
where: 

(a)
\be 
H_{hop}=-\sum\limits_{\left\langle {ij} \right\rangle } {t_{ij}c_{i\alpha }^+
c_{j\alpha }-\sum\limits_{\left[ {ij} \right]} {t'_{ij}}}c_{i\alpha }^+c_{j\alpha }-\sum\limits_{\left\{ {ij} \right\}} 
{t''_{ij}}c_{i\alpha }^+c_{j\alpha }\,,\label{2.2}\ee
and $\left\langle {\dots } \right\rangle $ 
denotes nearest neighbour (NN) sites, $\left[ {} \dots 
\right]$ next-to-nearest neighbour (NNN), and
$\left\{ {} \right\}$ third nearest neighbour.
Here repeated spin (or "colour") indices 
are summed over. The Latin indices $i,j$ 
denote lattice sites and the Greek indices $\alpha =1,2$
are spin components. 

(b) \be H_J=J\sum\limits_{\left\langle {ij} \right\rangle } {T_{i,\alpha \beta }}T_{j,\beta \alpha }+
J'\sum\limits_{\left[ {ij} \right]} 
{T_{i,\alpha \beta }}T_{j,\beta \alpha }\,,\label{2.3}\ee
with $T_{i,\alpha \beta }=c_{i\alpha }^+c_{i\beta }$.
The quantities $J,J'$ 
denote the couplings of the 
appropriate Heisenberg antiferromagnetic interactions.  
We shall be interested~\cite{[3]} in the regime where 
$J'<<J$.

(c) \be H_\mu =\mu \sum\limits_i {c_{i\alpha }^+c_{i\alpha }}\,,
\label{2.4}
\ee and $\mu$ is the chemical potential.

(d) \be H_V=-V\sum\limits_{\left\langle {ij} \right\rangle } {n_i}n_j\,,
\label{2.5a}
\ee
 and $n_i=\sum\limits_{\alpha =1}^2 {c_{i\alpha }^+
c_{i\alpha }}$. This is an effective static NN interaction which is provided in the bare $t-J$ model by the exchange term, 
because of the extra magnetic bond 
in the system 
when two polarons are on neighbouring sites~\cite{[1]}. In
ref. \cite{[1]} the strength of the interaction is taken to be:
\be V\approx 0.585\,J\,,\label{2.6a} \ee
This is related to the regime of the parameters used in \cite{[1]},
for which the NN hoping elements $t << J$. 
However, one may consider more general models~\cite{future},
in which the above restrictions 
are not valid, and $V$ is viewed as an independent 
parameter of the effective theory, e.g. 
\be V\approx b~\,J\,,\label{2.6ab} \ee
with $b$ a constant to be determined phenomenologically. 
As we shall discuss below, 
this turns out to be useful 
for the existence of 
supersymmetric points in the parameter space of the model. 

(e) The operator $P$ projects out double occupancy at a site.

We define the doping parameter $0 < \delta  < 1$ by 

\be \sum\limits_{\alpha =1}^2 {\left\langle {c_{i\alpha }^+c_{i\alpha }} \right\rangle }=1-\delta \,,\label{2.7}\ee

$d$-wave pairing, which seems to have been confirmed experimentally for high-$T_c$ cuprates, was assumed in 
\cite{[1]}. 
A d-wave gap is represented by an order parameter of the form
\be
\Delta \left( {\vec k} \right)
=\Delta _0\left( {\cos \,k_xa-\cos \,k_ya} \right) \,,\label{2.8b}
\ee
where $a$ is the lattice spacing.
The relevant Fermi surface is characterised by the following four 
nodes where the gap vanishes:
\be
\left( {\pm {\pi  \over {2a}},\pm {\pi  \over {2a}}} \right)\,,
\label{2.9b}\ee
We now consider the generalized dispersion relation \cite{[5],[6]} 
for the quasiparticles in the superconducting state:
\be
E\left( {\vec k} \right)=\sqrt {\left( {\varepsilon \left( {\vec k} \right)-\mu } \right)^2+\Delta ^2\left( {\vec k} \right)}\,,\label{2.10b}
\ee
In the vicinity of the nodes it is 
reasonable~\cite{[5],[6]} 
to assume that $\mu \approx 0$ or equivalently  we may
linearize about $\mu$, 
i.e. write $\varepsilon \left( {\vec k} \right)-\mu \approx v_D\left| {\vec q} \right|$~\cite{[2]} where
$v_D$ is the 
effective velocity at the node and $q$ is the wave-vector with 
respect to the nodal point.

\section{Non-Abelian spin-charge separation in the t-J model}

As already mentioned in the introduction, 
in ref. \cite{[5]} 
it was {\it proposed} that for the large-U limit of the {\it doped} 
Hubbard model the following {\it `particle-hole' symmetric 
spin-charge separation} 
ansatz occurs at {\it each site} $i$: 

\be\chi _{\alpha\beta,i} = 
\psi _{\alpha\gamma,i}z_{\gamma\beta,i} \equiv \left(
\begin{array}{cc}
c_1 \qquad c_2 \\
c_2^\dagger \qquad -c_1^\dagger \end{array} 
\right)_i = 
\left(\begin{array}{cc} 
\psi_1 \qquad \psi_2 \\
-\psi_2^\dagger \qquad \psi_1^\dagger\end{array} 
\right)_i~\left(\begin{array}{cc} z_1 \qquad -{\overline z}_2 \\
z_2 \qquad {\overline z}_1 \end{array} \right)_i \label{3.1}
\ee
where the fields $z_{\alpha,i}$ obey canonical {\it bosonic} 
commutation relations, and are associated with the
{\it spin} degrees of freedom (`spinons'),  
whilst the fields 
$\psi _{a,i},~a=1,2$ have {\it fermionic}
statistics, and are assumed to {\it create} 
{\it holes} at the site $i$ with spin index $\alpha$ (`holons'). 
The ansatz (\ref{3.1}) 
has spin-electric-charge separation, since only the 
fields $\psi$ carry {\it electric} charge. Generalization to the non-Abelian
model allows for inter-sublattice hopping of holes which is observed experimentally.

It is worth noticing that the anticommutation relations 
for the electron fields $c_\alpha$,$c_\beta^\dagger$,
do not quite follow from the ansatz (\ref{3.1}). 
Indeed, assuming the canonical 
(anti) commutation relations for the $z$ ($\psi$) fields, 
one obtains from the ansatz (\ref{3.1}) 
\bea
\{ c_{1,i}, c_{2,j} \} & \sim & 2 \psi_{1,i}\psi _{2,i}\delta_{ij} \nn \\
\{ c_{1,i}^\dagger, c_{2,j}^\dagger \} &\sim & 2 \psi_{2,i}^\dagger
\psi _{1,i}^\dagger \delta_{ij} \nn \\
\{ c_{1,i}, c_{2,j}^\dagger \} & \sim& \{ c_{2,i}, c_{1,j}^\dagger \}
\sim 0 \nn \\
\{ c_{\alpha,i}, c_{\alpha,j}^\dagger \} & \sim &  
\delta _{ij}\sum_{\beta =1,2} [z_{i,\beta} {\overline z}_{i,\beta} + 
\psi_{\beta,i}\psi _{\beta,i}^\dagger],~~\alpha=1,2
\qquad {\rm no~sum~over~i,j}\label{3.2}\eea
To ensure {\it canonical} commutation relations for the $c$ operators
therefore we must {\it impose} at each lattice site the 
(slave-fermion) constraints
\bea &~&\psi_{1,i}\psi_{2,i} 
= \psi^\dagger _{2,i}\psi^\dagger_{1,i} = 0,\nn \\
&~&\sum_{\beta =1,2} [z_{i,\beta} {\overline z}_{i,\beta} + 
\psi_{\beta,i}\psi _{\beta,i}^\dagger] = 1
\label{3.3}
\eea
Such relations are understood to be satisfied when the 
holon and spinon operators act on {\it physical} states.
Both of these relations are valid in the large-$U$ limit 
of the Hubbard model and encode the non-trivial physics 
of constraints behind the spin-charge separation ansatz 
(\ref{3.1}). They express the 
constraint {\it at most one electron or hole per site},
which characterizes the large-$U$ Hubbard models we are considering 
here. 

There is a local phase (gauge) non-Abelian symmetry hidden in the 
ansatz (\ref{3.1})~\cite{[5]}
$G=SU(2)\times U_S(1)$, where $SU(2)$ stems from 
the spin degrees of freedom, $U_S(1)$ is a 
statistics changing group, which 
is exclusive to two spatial dimensions and is responsible 
for transforming bosons into fermions and vice versa.
As remarked in \cite{[5]}, 
the $U_S(1)$ effective interaction is responsble for the 
equivalence between  the slave-fermion ansatz  
(i.e. where the holons are viewed as 
charged bosons and the spinons as electrically 
neutral fermions~\cite{[17]})
and the slave boson ansatz (i.e. where the holons are viewed as 
charged fermions and the spinons as neutral bosons~\cite{[18],[5]}). 
This is analogous (but not identical) to the bosonization 
approach of \cite{[19]} for anyon systems. 

The application of the ansatz (\ref{3.1}) 
to the Hubbard (or t-j models) 
necessitates a `particle-hole' symmetric formulation of the 
Hamiltonian (\ref{1.1}), 
which as shown in \cite{[5]}, 
is expressible
in terms of the operators $\chi$. 
Upon appropriate linearizations of the various four-field
operators involved using the Hubbard-Stratonovitch method,
we obtain 
the effective spin-charge separated action
for the doped-antiferromagnetic model of \cite{[5]}:
\bea
&~& H_{HF}=\sum_{<ij>} \left(tr\left[(8/J)\Delta^\dagger_{ij}\Delta_{ji}
+ \left| A_1 \right| (-t_{ij}(1 +
\sigma_3)+\Delta_{ij})\psi_j V_{ji}U_{ji}\psi_i^\dagger\right] + \right.\nn \\
&~& \left. tr\left[ K{\overline z}_iV_{ij}U_{ij}z_j\right] + h.c. \right)
+ \dots \,,
\label{1.6}
\eea
with the $\dots$ denoting 
chemical potential terms.
This form of the action, 
describes low-energy excitations 
about the Fermi surface of the theory. 
$\Delta _{ij}$ is a 
Hubbard-Stratonovich field that linearizes four-electron interaction
terms in the original Hubbard model.
The quantities
$V_{ij}$ and $U_{ij}$ denote lattice link variables associated with 
elements of the  $SU(2)$ and $U_S(1)$ groups respectively.
They are associated~\cite{[5]} 
with phases 
of vacuum expectation values 
of bilinears $<~{\overline z}_i z_j~>$ 
and/or $<~\psi^\dagger _i \left( -t_{ij}(1 + \sigma _3) + \Delta _{ij} \right)\psi_j~>$. 
It is understood that, 
by integrating out in a path integral over $z$ and $\psi$ variables,
fluctuations are incorporated, which go beyond a Hartree-Fock treatment. 
The quantity $\left|A_1 \right|$ is the amplitude of the bilinear 
$<~{\overline z}_i z_j~>$ assumed frozen~\cite{[5]}. 
By an appropriate normalization 
of the respective field variables, one may set $\left|A_1 \right| =1$,
without loss of generality. 
In this normalization, one may then 
parametrize the quantity $K$, which is the amplitude of 
the appropriate
fermionic bilinears, as~\cite{[5],[3]}:
\be
K \equiv \left(J \left| \Delta _z \right|^2  \eta^2 \right)^{1/2}~;
\qquad \eta \equiv <\sum_{\alpha =1}^{2}~\psi_\alpha \psi^\dagger_\alpha > 
=  1 - \delta~, \label{defK}
\ee
with $\delta $ the doping concentration in the sample.  
The quantity  
$|\Delta _z |$ is considered as an arbitrary 
parameter of our effective theory, of dimensions $[energy]^{1/2}$, 
whose magnitude  
is to be fixed 
by phenomenological or other considerations (see below). 
To a first approximation we assume that $\Delta _z $
is doping independent~\footnote{However, from its 
definition,  as a $< \dots >$ 
of a quantum model with complicated $\delta $ dependences in its 
couplings, the quantity $\Delta _z$ may indeed exhibit 
a doping dependence. For some consequences of this 
we refer the reader to the discussion in section 6, below, 
and in 
ref. \cite{future}.}.   
The dependence on $J $ and $\delta$ in (\ref{defK}) 
is dictated~\cite{[3]} by the
correspondence with the conventional antiferromagnetic
$CP^1$ $\sigma$-model in the limit $\delta \rightarrow 0$. 

The model of ref. \cite{[1]} differs from that of \cite{[5]}
in the existence of 
NNN hopping $t'$ and 
tripple neighbor hopping $t''$, which were ignored in the 
analysis of \cite{[5]}.  
For the purposes of this work, 
which focuses on the low-energy (infrared) 
properties
of the continuum field theory of (\ref{1.6}), 
this can be taken into 
acount 
by assuming that 
\be
|t_{ij}| = t'_{+} \equiv  t + 2t_+, 
\qquad t_+ \equiv t' + 2t''\label{2.6b}
\ee
in the notation of \cite{[1]}. 
The relation stems from the observation that 
in the continuum low-energy 
field-theory limit such 
$NNN$ and triplle hopping terms
can be Taylor expanded (in derivatives). 
It is the terms linear in derivatives that 
yield the shift (\ref{2.6b}) 
of the NN neighbor hopping element $t$.  
Higher derivatives terms, 
of the form $\partial _x \partial _y$ are suppressed 
in the low-energy (infrared) limit. 

It is important to note that the model of \cite{[5]},
as well as its extension (\ref{1.6}),  
in contrast to that discussed in \cite{[2]},  
involves only a {\it single lattice} structure, with nearest neighbor
hopping ($<ij>$) being taken into account, $t_{ij}$. 
The antiferromagnetic nature is then viewed as a `colour' degree of freedom,
being expressed via the non-Abelian gauge structure of the 
spin-charge separation ansatz (\ref{3.1}). 
As we shall discuss later, 
this is very important in yielding the correct number of fermionic 
(holons $\Psi$) 
degrees of freedom in the continuum low-energy field theory to match 
the bosonic degrees of freedom (spinons $z$) at the supersymmetric point. 

\section{The Effective Low-Energy Gauge Theory}

The conventional lattice gauge theory form of the action 
(\ref{1.6})
is derived upon
freezing the fluctuations of the $\Delta_{ij}$ 
field, 
assuming, as usual, the flux phase
for the gauge field $U_S(1)$, with flux $\pi$ per lattice plaquette,
and assembling the fermionic degrees of freedom into two 2-component
Dirac spinors~\cite{[5]}:
\be
{\tilde \Psi}_{1,i}^\dagger =\left(\psi_1~~-\psi_2^\dagger\right)_i,~~~~
{\tilde \Psi} _{2,i}^\dagger=\left(\psi_2~~\psi_1^\dagger\right)_i\label{2.5}
\ee
The fermionic part of 
the long-wavelegth lattice lagrangian, then, reads:
\bea
&~& S={1\over 2}K' \sum_{i,\mu}[{\overline \Psi}_i (-\gamma_\mu) 
U_{i,\mu}V_{i,\mu} \Psi_{i+\mu}  + \nn \\
&~& {\overline \Psi}_{i+\mu}
(\gamma _\mu)U^\dagger_{i,\mu}V^\dagger_{i,\mu}
\Psi _i ]  + {\rm Bosonic~CP^1~parts}\label{2.6}
\eea
where the Bosonic $CP^1$ parts denote magnon-field $z$ dependent 
terms, and  
are given in (\ref{1.6}). 
The coefficient $K'$ is a constant 
which stems from the $t_{ij}-$ and $\Delta _{ij}-$ dependent coefficients
in front of the fermion terms in (\ref{1.6}). 
The fermions $\Psi$ in (\ref{2.6}) are  
{\it two-component} `coloured' 
spinors,
related to  the spinors in (\ref{2.5}) 
via a Kawamoto-Smit transformation~\cite{[20]}
\be
\Psi _c(r) =\gamma_0^{r_0}\dots \gamma_2^{r_2}{\tilde \Psi} _c(r)
\qquad  {\overline \Psi} _c(r) 
={\overline {\tilde  \Psi }}_c(r)(\gamma_2^{\dagger})^{ r_2}\dots 
(\gamma_0^{\dagger})^{r_0}\label{2.8}\ee
where $r$ is a point on the spatial lattice,  
and $c$ is a  `colour'
index $c=1,2$ expressing the initial 
antiferromagnetic nature of the system; the 
$\gamma$ matrices are $2 \times 2$ antihermitean Dirac 
matrices on a Euclidean Lattice satisfying the 
algebra 
\be
\{ \gamma_\mu~,~\gamma _\nu \} =-2\delta _{\mu\nu}\label{2.9}\ee
In terms of the Pauli matrices $\sigma_i,i=1, \dots 3$, the 
$\gamma$ matrices are given 
by $\gamma _\mu = i\sigma_\mu,~\mu=1,2,3.$
Notice that 
fermion bilinears of the form  
${\overline \Psi}_{i,c} \Psi _{i,c'}$ ($i$=Lattice
index) 
are just 
\be
{\overline \Psi}_{i,c} \Psi _{i,c'} = 
{\overline {\tilde \Psi}}_{i,c} {\tilde \Psi}_{i,c'}\label{2.10}\ee
due to the 
Clifford algebra (\ref{2.9}), 
and (anti-) hermiticity properties 
of the $2\times 2$ $\gamma$ matrices
on the Euclidean lattice. On a lattice, in the path integral over
the fermionic degrees of freedom in a quantum theory, 
the variables ${\overline \Psi}$ and $\Psi$ are viewed as {\it independent}. 
In view of this, 
the spinors $\Psi^\dagger_\alpha $ 
in (\ref{2.5})
may be replaced by ${\overline \Psi _\alpha}$,     
as being path integral variables
on a Euclidean Lattice appropriate for the Hamiltonian 
system (\ref{1.1}). 
This should be kept in mind when discussing
the microscopic structure of the theory in terms of the 
holon creation and annihilation operators $\psi_\alpha^\dagger, \psi_\alpha,
\alpha=1,2$.

An order of magnitude estimate of the  
modulus of $\Delta _{ij}$ then, which determines the strength of the 
coefficient $K'$ may be provided by its equations of motion. 
Assuming that the modulus of (the dimensionless) 
fermionic bilinears is of order unity, then, we have as an order of magnitude 
\be
K' \sim \left(t'_{+} + {J \over 8}\right)
\label{anonymous}
\ee 
Notice that in the regime of the parameters of \cite{[1]} 
$t << t_+$ and $t_+ \simeq {3 \over 2}J$ 
for a momentum regime close to a node in the fermi surface, 
of interest to us here. Thus 
\be
K' \simeq 25J/8 \label{2.6c}\ee 
For reasons that will become clear below
we may consider a regime of the parameters of the theory for which 
\be
K' >> K = \sqrt{J}\left|\Delta _z \right|\left(1 - \delta \right),\qquad 
0 < \delta < 1   
\label{2.6d}\ee 
For the model of \cite{[1]}, for instance, 
on account of (\ref{2.6c}), this condition implies
that 
\be
\sqrt{J}/\left|\Delta _z \right| \gg 0.32~(1-\delta)~,\qquad 
0 < \delta < 1   
\label{cond2}
\ee
By 
appropriately rescaling the fermion fields $\Psi$ to $\Psi'$, so that in the continuum they have a canonical Dirac term, 
we may 
effectively constrain
the $z$ fields to satisfy the 
$CP^1$ constraint:
$$ |z_\alpha |^2 + {1\over K'} ({\rm \Psi'-bilinear~terms}) 
=  1$$
where now the fields $\Psi$ are dimensionful. 
with dimensions of 
$[energy]$. A natural order of magnitude of these dimensionful 
fermion bilinear 
terms is of the order of $K^2$, which plays the r\^ole of the characteristic 
scale in the theory, being related directly to the 
Heisenberg exchange energy $J$.  
In the limit $K' >> K$ (\ref{2.6d}) therefore  the fermionic terms in the 
constraint can be  ignored, and the constraint assumes the standard 
$CP^1$ form involving only the $z$ fields ( this being also the case 
for the model of \cite{[2],[3]},
in a specific regime of the 
microscopic parameters).  
As discussed in \cite{[8],future}, 
such a form for the constraint is the one  
{\it appropriate for supersymmetrization}. 

As we shall see later, however, 
the condition (\ref{cond2}) alone, although 
necessary, is not sufficient to guarantee 
the existence of supersymmetric points.  
Supersymmetry imposes 
additional restrictions,  
which in fact rule out 
the existence of supersymmetric points for the 
model of \cite{[1]} compatible with 
superconductivity~\footnote{We note in 
passing that in realistic materials superconductivity 
occurs for doping concentrations above $3\%$, and is destroyed 
for doping concentrations larger than $\delta _{\rm {max}} \sim 10\%$.}. 
However, this does not prevent 
one from considering more general models~\cite{future} 
in which 
$K'$ is viewed as a phenomenological parameter, not constrained 
by (\ref{2.6c}). In that case, supersymmetric points may occur 
for a certain regime of the respective parameters.

With the above in mind we consider from now on the standard 
$CP^1$ constraint involving only $z$ fields. 
By an appropriate normalization of $z$ to $z' = {z \over \sqrt{1 - \delta}}$
the constraint then acquires the familiar normalized $CP^1$ form 
$|z_\alpha|^2=1$ form.
This implies a rescaling of the normalization coefficient $K$  
in (\ref{1.6}):  
\be
K \rightarrow  {1 \over \gamma} \equiv K (1-\delta) \simeq \sqrt{J}|\Delta _z| (1-\delta)^2  
\label{3.6e}
\ee
In the naive continuum limit, then, the effective lagrangian 
of spin and charge degrees of freedom describing the low-energy 
dynamics of the Hubbard (or $t-j$) model (\ref{1.6}) of \cite{[5]} 
is then: 
\be
{\cal L}_2 \equiv  {1 \over \gamma }{\rm Tr}\left| \left( {\partial _\mu +
ig \sigma^a B_\mu^a + iga_\mu } \right)z \right|^2 +
{\overline \Psi}D_\mu\gamma_\mu\Psi \label{3.7}\ee
with $z_\alpha$ a complex doublet satisfying the constraint 
\be
|z_\alpha|^2=1\label{3.7b}
\ee
The Trace ${\rm Tr}$ is over group indices, 
$D_\mu = \partial_\mu -ig_1a_\mu^S-ig_2\sigma^aB_{a,\mu}
-{e \over c}A_\mu$,
$B_\mu^a$ is the gauge potential of the local (`spin') $SU(2)$ group,
and $a_\mu$ is the potential of the $U_S(1)$ group.

It should be remarked that, 
we are 
working in units of the Fermi velocity $v_F (=v_D)$ of holes, which plays the 
r\^ole of the limiting velocity for the nodal liquid. 
For the nodal liquid at the supersymmetric points 
we also assume that $v_F \simeq v_S$, where $v_S$ is the 
effective velocity of the spin degrees of freedom. 
The relativistic form of the fermionic and bosonic
terms of the action (\ref{3.7}) 
is valid {\it only} in this regime
of velocities. This is sufficient for our purposes in this work. 
Indeed, at the supersymmetric points, 
where we shall restrict our analysis here, 
the mass gaps 
for spinons and holons, which may be generated {\it dynamically},  
are {\it equal} by virtue of supersymmetry at zero temepratures and
in the absence of any external fields. Hence it makes sense
to assume the equality in the propagation velocities
for spin and charge degrees of freedom,
given that this situation is consistent with the 
respective dispersion relations. This is {\it not true},
of course, for excitations away from such points.

\section{The NN interaction terms $H_V$ }

We will now discuss the terms 
\be
H_V=-V\sum\limits_{\left\langle {ij} \right\rangle } {n_in_j}
\label{anon2}
\ee
introduced in ref. \cite{[1]}. 
With the above discussion in mind 
for the spinors (\ref{2.5})
we note that, under the ansatz (\ref{3.1}), at a site $i$
the electron number operator  $\eta_i$ 
is expressed, through the Determinant (Det) of the $\chi$ matrix in 
(\ref{3.1}),  
in terms of the spin, $z_\alpha,\alpha=1,2$, 
and charge $\psi_\alpha, \alpha=1,2$, operators 
as: 
\bea
&~&\eta_i \equiv \sum_{\alpha = 1}^{2}
 c^{\dagger}_{\alpha,i}c_{\alpha,i}={\rm Det}\chi_{\alpha\beta,i} = \nn \\
&~&{\rm Det}{\hat z}_{\alpha\beta,i} + {\rm Det}{\hat \psi}_{\alpha\beta,i} =
\sum_{\alpha=1}^2 \left( \psi_\alpha 
\psi_\alpha ^\dagger + |z_\alpha|^2 \right)
\label{4.1}\eea 
We may express the quantum fluctuations for 
the Grassman fields  
$\psi_\alpha$ (which now carry a `colour' index $\alpha=1,2$
in contrast to Abelian spin-charge separation models) 
via: 
\be
\psi _{\alpha, i}\psi _{\alpha,i}^+=\left\langle {\psi _{\alpha, i}
\psi _{\alpha, i}^+} \right\rangle +  
:\psi _{\alpha, i}\psi _{\alpha, i}^+:\,,~{\rm no~sum~over~i}
\label{4.2}\ee
where $: \dots :$ denotes normal ordering of quantum operators,
and from now on, unless explicitly stated, 
repeated indices are summed over. 
Since $$\left\langle {\psi _{\alpha, i}\psi _{\alpha, i}^+} 
\right\rangle \equiv 1-\delta~,~~~{\rm no~sum~over~i}$$ 
$\delta$ the doping concentration 
in the sample (\ref{2.7}),  
we may rewrite $\eta_i$ as $$\eta_i = |z_\alpha |^2 + (1 -\delta) 
+ :\psi_\alpha \psi_\alpha^\dagger: $$
which in terms of the spinors $\Psi$ 
is given by (c.f. 
(\ref{2.5}),(\ref{2.10})): 
\be
\eta_i = 2 -\delta + {1 \over 2} 
\left(\Psi^\dagger_\alpha \sigma_3 
\Psi_\alpha \right)_i\label{3.3b}\ee
where $\sigma_3 = \left(\begin{array}{cc}
1 \qquad 0 \nn \\
0 \qquad -1 \end{array}\right)$ acts in (space-time) 
spinor space,
and we took  
into account the $CP^1$ constraint (\ref{3.7b}).  

Consider now the attractive interaction term $H_V$ (\ref{anon2}), 
introduced 
in ref. \cite{[1]}. 
We then observe than 
the terms linear in $(2-\delta)$ in the expression for $H_V$  
can be absorbed by an appropriate shift 
in the chemical potential, about which we linearize to obtain the 
low-energy theory. We can therefore ignore such terms from now on.  
 
Next, we make use of the fact, mentioned earlier, 
that in a Lattice path integral 
the spinors $\Psi^\dagger_\alpha $ 
may be replaced by ${\overline \Psi _\alpha}$.    
From the structure of the spinors (\ref{2.5}), then,  
we observe that we may rewrite the $H_V$ term 
{\it effectively} as
a Thirring vector-vector interaction among the spinors $\Psi$ 
\be
H_V=+{V \over 4} \sum_{<ij>} \left( {\overline \Psi}_\alpha \gamma _\mu \Psi _\alpha \right)_i
\left( {\overline \Psi}_\beta \gamma ^\mu \Psi _\beta \right)_j \label{4.5}\ee
where summation over the repeated indices $\alpha, \beta (=1,2)$, 
and $\mu=0,1,2$, with $\mu=0$ 
a temporal index,
is understood. To arrive 
at (\ref{4.5}) we have expressed $\sigma_3$ as $-i\gamma_0$, and  
used the Clifford algerba (\ref{2.9}),
the off-diagonal nature of the $\gamma_{1,2}=i\sigma_{1,2}$ matrices,
as well as the constraints
(\ref{3.3}). In particular the latter imply that any scalar product 
between Grassmann variables $\psi_\alpha$ (or $\psi^\dagger_\beta$)  
with different `colour' indices {\it vanish}. 

Taking the  continuum limit of (\ref{4.5}), and ignoring 
higher derivative terms involving four-fermion 
interactions, which by power counting are irrelevant 
operators in the infrared, 
we obtain after passing to a Lagrangian formalism
\be
{\cal L}_V = 
-{V \over 4 K'^2} \left( {\overline \Psi _\alpha} \gamma _\mu \Psi _\alpha
\right)^2 \label{4.6}\ee
where we have used rescaled spinors, with 
the canonical Dirac kinetic term with unit coefficient,  
for which the canonical form of the $CP^1$ 
constraint (\ref{3.7b}) 
is satisfied. For notational convenience 
we use the same notation $\Psi$ for these spinors as the unscaled ones. Although this 
is called the naive continuum limit, it captures correctly the leading infrared behaviour
of the model.

We then use a Fierz rearrangement formula for the $\gamma$ matrices  
$$\gamma ^\mu _{ab}\gamma _{\mu,cd} = 
2 \delta _{ad}\delta _{bc} - \delta _{ab}\delta _{cd}$$
where Latin letters indicate spinor indices, and Greek Letters
space time indices.  
The Thirring (four-fermion) 
interactions (\ref{4.5}) then 
become: 
\be
\left({\overline \Psi}_\alpha \gamma _{\mu} \Psi _\alpha \right)^2= 
-3 \left({\overline \Psi} _\alpha \Psi _\alpha \right)^2 
- 4 \sum_{\alpha < \beta} 
\left({\overline \Psi }_\alpha \Psi _\beta  {\overline \Psi }_\beta 
\Psi _\alpha \right)  
\label{4.7}\ee
As mentioned above, 
in the model of \cite{[5]}, due to the first of the constraints
(\ref{3.3}), the mixed colour terms vanish, thereby leaving us 
with pure Gross-Neveu {\it attractive} interaction terms of the form: 
\be
{\cal L}_V = +{3 V \over 4 K'^2} \left({\overline \Psi} _\alpha \Psi _\alpha \right)^2
\label{3.8}\ee
which describe the low-energy 
dynamics of the interaction (\ref{anon2}) in the context 
of the non-Abelian spin-charge separation (\ref{3.1}). 
It should be
stressed that (\ref{3.8}) is specific to our spin-charge 
separation model.

Moreover in the context of the spinors (\ref{2.5}), a 
condensate of the form $<{\overline \Psi}_\alpha \Psi_\alpha>$ 
on the lattice {\it vanishes}
because of the constraints (\ref{3.3}). 
Such condensates would violate parity (reflection) 
operation on the planar spatial lattice, which on the spinors ${\tilde \Psi}$ 
is defined to act as follows: 
$${\tilde \Psi}_1 \left( x \right) \rightarrow \sigma_1 {\tilde \Psi}_2\left( x \right),\quad 
{\tilde \Psi}_2\left( x \right) \rightarrow \sigma_1 {\tilde \Psi}_1\left( x \right)$$
or equivalently, 
in terms of the (microscopic) holon operxtors $\psi_\alpha, \alpha=1,2,$:
$$\psi_1\left( x \right) \rightarrow \psi_2^\dagger\left( x \right), \qquad 
\psi_2\left( x \right) \rightarrow -\psi_1^\dagger\left( x \right).$$   

To capture correctly this fact in the context of our effective continuum 
Gross-Neveu interaction (\ref{3.8}) 
the coupling strength {\it must} be subcritical,
i.e. weaker than the critical coupling for mass generation.  
The critical coupling of the Gross-Neveu interaction is 
expressed in terms of a high-energy cut-off scale $\Lambda$ as~\cite{[9]}: 
\be
1=4 g_c^2 \int\limits_{S_\Lambda } {{{d^3q} \over {8\pi^3 q^2}}} = {2 g_c^2 \Lambda \over \pi ^2}
\label{3.9}
\ee
where $q$ is a momentum 
variable and ${S_\Lambda }$ is a sphere of radius $\Lambda$. The divergent $q$-integral is cut-off at a momentum scale $\Lambda$ 
which defines the low-energy theory of interest. 
For the case of interest $g^2={3 V \over 4 K'^2}$; on 
using (\ref{2.6c}), then, the condition of sub-criticality requires that
\be
\Lambda  < 10^2~J~.
\label{3.10}
\ee 
which is in agreement with the fact that in all  
effective models for doped antiferromagnets used in the 
literature
the Heisenberg exchange energy serves as
an upper bound for the energies of the 
excitations
of the effective (continuum) theory. 

\section{Conditions for N=1 Supersymmetry and Potential 
Phenomenological Implications} 

We turn now to conditions for supersymmetrization of the 
above continuum theory. 
Below we shall sketch only the main results, which will 
be sufficient for the purposes of this talk.
Details will appear in a forthcoming publication~\cite{future}.
For simplicity we
shall ignore the non-Abelian $SU(2)$ interactions, keeping only the 
Abelian $U_S(1)$ ones, which has been shown 
to be responsible for dynamical 
mass generation (and superconductivity) in the model of \cite{[5]}. 
The extension to supersymmetrizing the full gauge multiplet 
$SU(2) \times U_S(1)$ will be the topic of 
a forthcoming work. However we shall still maintain 
the colour structure in the spinors, which is important 
for the ansatz (\ref{3.1})~\footnote{Ignoring the 
$SU(2)$ interactions implies, of course, that the `colour' structure
becomes a `flavour' index; however, this is essential  
for keeping track of the correct degrees of freedom 
required by supersymmetry in the problem at hand~\cite{[8]}.}. 

As discussed in detail in \cite{[8],[11]} 
the conditions for $N=1$ 
supersymmetric extensions of a $CP^1$ $\sigma$ model
is that 
the constraint is of the standard $CP^1$ form (\ref{3.7b}), 
supplemented by {\it attractive} 
four-fermion interactions of the Gross-Neveu type 
(\ref{3.8}), whose coupling is related to the 
coupling constant of the kinetic $z$-magnon terms of the 
$\sigma$-model in a way so as to guaranteee the balance 
between bosonic and fermionic degrees of freedom 
Specifically, 
in terms of component fields, the pertinent lagrangian reads:
\be
L = g_1^2 [D_{\mu} \bar{z}^{\alpha}D^{\mu}z^{\alpha} +
i {\overline \Psi} \not{D} \Psi + \bar{F}^{\alpha} F^{\alpha} 
 + 2i({\overline \eta} \Psi^{\alpha} \bar{z}^{\alpha} -
{\overline \Psi}^{\alpha} \eta z^{\alpha})]
\label{B.16a} 
\ee
where $D_\mu$ denotes the gauge covariant derivative with respect to the 
$U_S(1)$ field. 
The analysis of \cite{[8],[11]} shows that 
\be
{\overline F}^\alpha F_\alpha = \sum_{\alpha=1}^{2} {1 \over 4} \left( 
{\overline \Psi}^\alpha \Psi_{\alpha} \right)^2 
\label{B.19a}
\ee
We thus observe that the $N=1$ supersymmetric extension 
of the $CP^1$ $\sigma$ model {\it necessitates} 
the presence of {\it attractive} Gross-Neveu type interactions among the 
Dirac fermions of {\it each sublattice}, 
in addition to the gauge interactions. 

In the context of the effective theory (\ref{3.7}), (\ref{4.6}), 
discussed 
in this article,
the $N=1$ supersymmetric
effective lagrangian (\ref{B.16a}) is obtained 
under the following restrictions among the 
coupling constants of the statistical model: 
\be
{3 V \over K'^2} = \gamma = {1 \over \sqrt{J} |\Delta _z| 
(1 -\delta )^2}~,\qquad 0 < \delta  < 1    \label{4.1b}\ee
Note that in the context of the model of 
ref. \cite{[1]},  
for which (\ref{2.6a}),(\ref{2.6c}) are valid,
the relation 
(\ref{4.1b}) 
gives the supersymmetric point 
in the parameter space of the model 
at the particular
doping concentration $\delta = \delta _s$:
\be
\left(1 - \delta _s\right )^2 \simeq {5.56 \sqrt{J} \over |\Delta _z|}~,
\qquad 0 < \delta _s < 1   
\label{4.2b}
\ee
Then, compatibility with (\ref{2.6d}),(\ref{cond2})
requires : 
$1-\delta_s \gg 1.8$, which 
implies that the model of \cite{[1]}
does not have supersymmetric points. 
However, 
one may consider 
more general models~\cite{future} 
in which $V$ and $K' \sim
t'_+ + J/8 $ 
are treated as independent phenomenological parameters
(c.f. (\ref{2.6ab})); in such a case
one can obtain regions of parameters
that characterize the supersymmetric points (\ref{2.6d}),(\ref{4.1b}), 
compatible 
with superconductivity. 

Some comments are now in order:

First, it is quite important to remark that 
in the model of \cite{[5]}, where the 
antiferromagnetic structure of the theory is encoded in a colour 
(non-Abelian) degree of freedom of the spin-charge separated 
composite electron operator (\ref{1.5}) on a single lattice geometry,  
there is a matching between the 
 bosonic ($z$ spinon fields) and fermionic ($\Psi$ holon fields)  
physical degrees of freedom, as required by supersymmetry,         
without the need for 
duplicating them by introducing ``unphysical'' degrees 
of freedom~\cite{[8]}. 

The gauge multiplet of the $CP^1$ $\sigma$-model also needs a supersymmetric
partner which is a Majorana fermion called the gaugino.  
As shown in \cite{[8]}, such terms lead to an 
effective electric-charge violating interactions on the spatial planes,
given that the Majorana gaugino is a real field, and as such cannot
carry electric charge (which couples as a phase to a Dirac field). These
terms can be interpreted as the removal or addition of electrons due to
interlayer hopping.   

Another important point we wish to make concerns the 
four-fermion attractive Gross-Neveu interactions in 
(\ref{B.16a}),(\ref{B.19a}). 
As discussed in detail in \cite{fmm,future},
if the coupling of such terms is supercritical, then a parity-violating 
fermion (holon) mass would be generated in the model.  
However, the condition (\ref{3.10}), which is valid in the 
statistical model of interest to us here, implies that the 
respective coupling is always subcritical, and thus there 
is no parity-violating dynamical mass gap for the holons, induced 
by the contact  Gross-Neveu interactions.  This leaves one with the 
possibility of {\it parity conserving} dynamical mass generation,
due to the statistical gauge interactions in the model~\cite{[5],fmm}. 

A detailed analysis of such phenomena in the context
of our $CP^1$ model is left for future 
work. We note at present, however, that 
in $N=1$ supersymmetric gauge models, supersymmetry-preserving 
dynamical mass is possible~\cite{[8],cm,cmp}. In fact,
as discussed in \cite{cmp}, although by supersymmetry 
the potential is zero, and thus there would naively seem that there is 
no obvious 
way of selecting the non-zero mass ground state
over the zero mass one, however there appear to be instabilities
in the {\it quantum effective action} in the massless phase, 
which manifest themsleves through 
instabilities of the pertinent running coupling.  

From a physical point of view, such a phenomenon would imply 
that, for sufficiently strong gauge couplings, 
the zero temperature liquid of excitations 
at the nodes of a $d$ wave superconducting gap would be characterized  
by the dynamical opening of mass gaps for the holons. 
At zero temperature, and for the specific doping 
concentrations corresponding to the supersymmetric 
points, as advocated above, the nodal gaps between spinon and holons 
would be equal, in agreement with the assumed equality 
of the respective propagation velocities $v_F=v_S$,
which yielded the relativistic form of the effective continuum 
action (\ref{3.7}) of the nodal excitations
at the supersymmetric points. 
Moreover, it is known~\cite{[2],[3],[5]} that in the context of the
gauge model,  
under the influence of 
an external electromagnetic field, the nodal gap may become 
{\it superconducting}, with a Kosterlitz-Thouless (KT) 
type superconductivity,
not characterized by a local order parameter. 

At finite temperatures, however, 
at which supersymmetry 
is explicitly broken, this equality of mass gaps 
would disappear. Moreover,  
as the crude analysis of \cite{fmijmpb} indicates, 
such gaps would disappear at temperatures which are much lower than the 
critical temperature of the (bulk) $d$-wave superconducting gap. 
For instance, for a typical set of the parameters of the $t-j$ model
used in \cite{fmijmpb},
the nodal critical temperature is of order of a few $mK$, which is much 
smaller than the $100$ K bulk critical temperature    
of the high temperature superconductors. 
The application of an external magnetic field in the strongly type II,
high-temperautre superconducting oxides,
which is another source for explicit breaking of the 
potential supersymmetry, 
enhances the critical 
temperature~\cite{fmijmpb}
up to $30$ K, thereby providing a potential explanation 
for the 
recent findings of \cite{ong}.

However, if such situations with broken supersymmetry are viewed 
as cases of perturbed supersymmetric points, then one might 
hope of obtaining non-perturbative information on the 
phase structure of the liquid 
of nodal excitations in spin-charge 
separating scenaria of (gauge) 
high-temperature superconductors. This 
may also prove useful for a complete physical 
understanidng of the 
entire phenomenon, including excitations 
away from the nodes. 
In fact, as discussed 
in detail in \cite{fmm}, the presence of supersymmetric points
at certain doping concentrations, would favour superconductivity
due to the suppression of potentially dangerous non-perturbative effects
(instantons) of the compact statistical 
gauge field that would be 
responsible for giving   
the gauge field a mass, thereby destroying the superconducting 
nature of the gap. 
In the model of \cite{[5]} such instanton 
configurations 
are unavoidable due to the non-Abelian nature of the 
gauge symmetry characterizing the 
spin-charge separation 
ansatz (\ref{1.5}). In \cite{fmm} a breakdown of 
superconductivity due to instanton effects 
has been interpreted as 
implying a ``pseudogap'' phase: a phase in which 
there is dynamical generation of a mass gap for the 
nodal holons, which, however, is not characterized 
by 
superconducting properties.

\begin{figure}[htb]
\epsfxsize=4in
\bigskip
\centerline{\epsffile{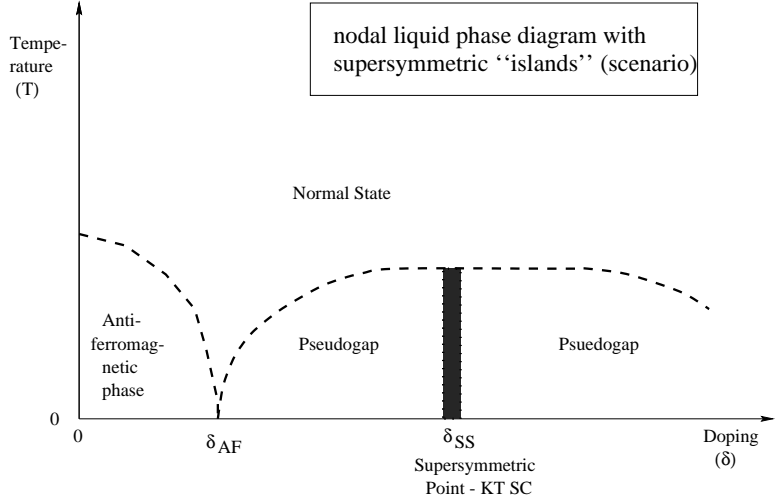}}
\vspace{0.2in}
\caption{\it A possible scenario for the 
temperature-doping phase diagram 
of a charged, relativistic, nodal liquid 
in the context of spin-charge separation.
At certain doping concentrations ($\delta_{SS}$) 
there are
dynamical supersymmetries among the spinon and holon 
degrees of freedom, responsible for yielding   
thin ``stripes'' in the phase diagram (shaded region)
characterized by 
Kosterlitz-Thouless (KT) superconductivity
without a local order parameter. The diagram 
is conjectural at present. It pertains strictly
to the nodal liquid excitations about the 
d-wave nodes of a superconducting gap, and 
hence,  
should not 
be confused with the phase diagram of the entire 
(high-temperature) superconductor.} 
\label{nodalfigure} 
\end{figure}

In this respect,  
the supersymmetric points (\ref{2.6d}),(\ref{4.1b}),
for which such instanton effects are 
argued~\cite{fmm} to be strongly suppressed in favour 
of KT superconductivity, 
would constitute 
``superconducting stripes'' in the temperature-doping 
phase diagram 
of the nodal liquid 
(see fig. \ref{nodalfigure})~\footnote{It should be stressed
that the term ``stripe'' here is meant to denote 
a certain region of the temperature-doping phase 
diagram 
of the nodal liquid and should not be confused 
with the stripe structures in real space which 
characterizes the cuprates at special doping concentrations.}. 
Theoretically, the stripes should have zero thickness,
given that they occur for specific doping concentrations
(\ref{4.1b}). However, in practice, there may be uncertainties
(due to doping dependences) in the precise value 
for the parameter $\Delta _z$ entering (\ref{4.1b}), 
which might be 
responsible for giving the superconducting stripe a
certain (small) thickness. 
A detailed analysis of such important issues
is still pending. It is hoped that due to supersymemtry 
one should be able to discuss some exact analytic 
results
at least for zero temperatures. 

\section{Conclusions} 

{}From the above discussion it is clear 
that supersymmetry 
can be achieved in 
the effective continuum field theories of 
doped antiferromagnetic systems 
exhibiting spin-charge separation 
only for {\it particular doping concentrations} 
(cf. (\ref{2.6d}),(\ref{4.1b})). 
One's hope is that the ancestor lattice model
will lie in the same universality class (in the infrared) 
as the continuum model, 
in the sense that it differs from it only by the action of 
renormalization-group irrelevant operators. This remains to be checked
by explicit lattice calculations. We should note at this stage
that this is a very difficult problem; in the context
of four-dimensional particle-physics models it is still 
unresolved~\cite{lattice}. 
However, in view of the apparent simpler form of the three-dimensional 
lattice models at hand, one may hope that these 
models are easier to handle.

By varying the doping concentration in the 
sample, one goes away from the supersymmetric point and 
breaks supersymmetry explicitly at zero temperatures. At finite 
temperatures,
or under 
the influence of external 
electromagnetic fields at the nodes of the d-wave gap,  
supersymmetry 
will also be broken explicitly. 
Therefore, realistic systems observed in nature will 
be characterized by explicitly broken supersymmetries
even close to zero temperatures. 
However there is value in deriving such supersymmetric results
in that at such points 
in the parameter space of the condensed-matter system
it is possible to obtain analytically 
some exact results on the phase structure of the theory. 
Supersymmetry may allow for a study of the 
quantum fluctuations about some 
exact ground states of the spin-charge separated
systems in a controlled way. 
Then one may consider perturbing around such exact solutions 
to get useful information about the non-supersymmetric models.

We have argued 
that such special 
points will yield KT superconducting ``islands''
in a temperature doping phase diagram 
of the nodal liquid, upon the dynamical 
generation of holon-spinon mass gaps (of equal size). 
This is 
due to special properties of the supersymmetry, 
associated
with the suppression of non-perturbative effects of the 
(compact) gauge fields entering the spin-charge 
separation ansatz (\ref{1.5}).  
This, of course, needs to be checked 
explicitly by carrying out the appropriate instanton 
calculations in the spirit 
of the non-perturbative modern framework of \cite{[15]}. 
At present, such non-perturbative effects 
can only be checked explicitly in three dimensions 
for highly extended supersymmetric models~\cite{dorey}.
It is, however, possible that
some exact results could be obtained at least for 
the $N=2$ 
supersymmetric 
models which may have some relevance for the
effective theory of the nodal liquid at the supersymemtric 
points~\cite{[8]}. Then, one may get 
some useful
information for the 
$N=1$ models by viewing the latter as 
supersymmetry-breaking perturbations  
of the $N=2$ models.
Such issues remain for future investigations,
but we hope that the speculations 
made in the present 
work provide 
sufficient motivation to carry out 
research along these directions.

\par

\par\noindent
\section*{Acknowledgements}\par\noindent
We thank the organizers and participants
of the TMR meeting {\it Topology and Phase Transitions
in Hot Matter}, 19-26 September 1999, Corfu (Greece), 
and of the adjacent Workshop {\it Common Trends in 
Particle and Condensed Matter Physics 1999}, 24-28 September 1999, 
where this 
work was presented, 
for their interest and advice. 
We also wish to thank I.J.R.~Aitchison, K.~Farakos, 
J.H.~Jefferson
and G.~Koutsoumbas for discussions.  
The work of N.E.M. is partially supported by
a P.P.A.R.C. (U.K.) Advanced Fellowship.

\end{document}